\documentclass[letterpaper, 10 pt, conference]{ieeeconf}  

\IEEEoverridecommandlockouts                              

\usepackage{cite}
\usepackage{amsmath,amssymb,amsfonts}
\usepackage{algorithmic}
\usepackage{graphicx}
\usepackage{textcomp}
\usepackage{xcolor}
\usepackage{caption}
\usepackage{multicol}
\usepackage{subcaption}

\usepackage{amsthm}
\usepackage[hidelinks]{hyperref}
\usepackage{float}
\usepackage[acronym]{glossaries}

\theoremstyle{definition}
\newtheorem{definition}{Definition}

\theoremstyle{plain}
\newtheorem{theorem}{Theorem}

\theoremstyle{plain}
\newtheorem{problem}{Problem}

\theoremstyle{remark}
\newtheorem{remark}{Remark}

\title{\LARGE \bf
Simultaneous Synthesis and Verification of Neural Control Barrier Functions through Branch-and-Bound Verification-in-the-loop Training
}

\author{Xinyu Wang$^{1}$,  Luzia Knoedler$^{1}$,  Frederik Baymler Mathiesen$^{2}$, and  Javier Alonso-Mora$^{1}$ 
\thanks{This paper has received funding from the European Union’s Horizon 2020 research and innovation programme under grant agreement No. 101017008.
Views and opinions expressed are however those of the author(s) only and do not necessarily reflect those of the European
Union or the European Research Council Executive Agency.
Neither the European Union nor the granting authority can be held responsible for them.}
\thanks{$^{1}$Xinyu Wang, Luzia Knoedler, and Javier Alonso-Mora are with the Cognitive Robotics Department,
        Delft University of Technology, 2628 CD Delft, The Netherlands
        {\tt\small x.wang-55@student.tudelft.nl}
        {\tt\small \{l.knoedler, j.alonsomora\}@tudelft.nl}}%
\thanks{$^{2}$Frederik Baymler Mathiesen is with the Delft
Center for Systems and Control, Delft University of Technology, 2628 CD Delft, The Netherlands
        {\tt\small f.b.mathiesen@tudelft.nl}}%
}

\newacronym{cbf}{CBF}{Control Barrier Function}

\newacronym{bf}{BF}{Barrier Function}

\newacronym{nn}{NN}{Neural Network}

\newacronym{fcnn}{FCNN}{Fully-Connected Neural Network}

\newacronym{ncbf}{nCBF}{Neural Control Barrier Function}

\newacronym{ce}{CE}{Counterexample}

\newacronym{cbvf-vi}{CBVF-VI}{Control Barrier-Value
Function Variational Inequality}

\newacronym{bbv}{BBV}{Branch-and-Bound Verification scheme}

\newacronym{bbs}{BBS}{Branch-and-Bound scheme}

\newacronym{bbvt}{BBVT}{Branch-and-Bound Verification-in-the-loop Training}

\newacronym{smt}{SMT}{Satisﬁability
Modulo Theory}

\newacronym{qp}{QP}{Quadratic Program}

\newacronym{sos}{SOS}{sum-of-squares}

\newacronym{rl}{RL}{Reinforcement Learning}

\newacronym{cbvf}{CBVF}{Control Barrier-Value Function}

\newacronym{hji-ra}{HJI-RA}{Hamilton-Jacobian-Issac Reachability Analysis}

\newacronym{sgd}{SGD}{Stochastic Gradient Descent}

\newacronym{nlp}{NLP}{Nonlinear Program}

\begin{document}

\maketitle
\thispagestyle{empty}
\pagestyle{empty}

\begin{abstract}

\acrfullpl{cbf} that provide formal safety guarantees have been widely used for safety-critical systems. However, it is non-trivial to design a \acrshort{cbf}. Utilizing neural networks as \acrshortpl{cbf} has shown great success, but it necessitates their certification as \acrshortpl{cbf}. In this work, we leverage bound propagation techniques and the \acrlong{bbs} to efficiently verify that a neural network satisfies the conditions to be a \acrshort{cbf} over the continuous state space. To accelerate training, we further present a framework that embeds the
verification scheme into the training loop to synthesize and verify
a neural \acrshort{cbf} simultaneously.
In particular, we employ the verification scheme to identify partitions of the state space that are not guaranteed to satisfy the \acrshort{cbf} conditions and expand the training dataset by incorporating additional data from these partitions.
The neural network is then optimized using the augmented dataset to meet the \acrshort{cbf} conditions.
We show that for a non-linear control-affine system, our framework can efficiently certify a neural network as a \acrshort{cbf} and render a larger safe set than state-of-the-art neural \acrshort{cbf} works. We further employ our learned neural \acrshort{cbf} to derive a safe controller to illustrate the practical use of our framework.
\end{abstract}

\section{Introduction}

Safety is a critical element of autonomous systems, such as self-driving cars and manipulators that interact with humans. As autonomous systems grow more complex, determining whether they operate safely becomes challenging. 

\begin{figure}
    \centering
    \includegraphics[width=0.5\textwidth]{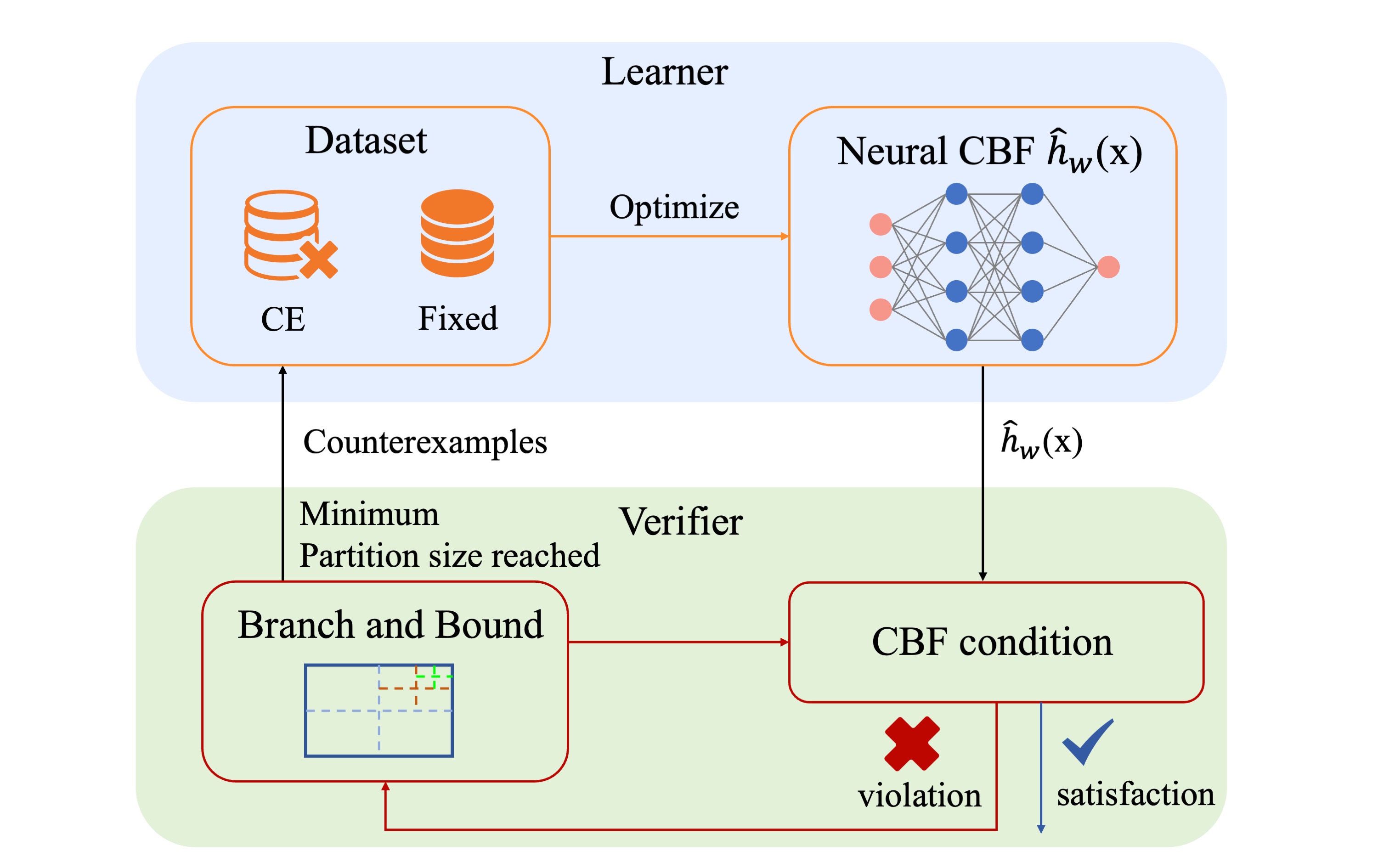}
    \caption{A schematic overview of the presented \acrlong{bbvt}. 
    The framework comprises of two key components: the learner and the verifier, which operate sequentially. The learner optimizes the \acrshort{ncbf} using a fixed dataset and a \acrlong{ce} dataset.
    The verifier leverages bound propagation techniques and the \acrlong{bbs} to refine a partition of the state space until the \acrshort{cbf} conditions are satisfied or counterexamples are generated. }
    \label{fig: system diagram}
\vspace{-10pt}
\end{figure}

Safety can be formulated via invariance, in the sense that any trajectory originating within an invariant set will never traverse beyond the boundaries of that set.
Lately, the use of \glspl{cbf} to derive a forward invariant set has received significant attention in the control and learning community \cite{dawson2023safe}. 
However, there exists no general and scalable technique for designing \glspl{cbf}. 
Therefore, recent works \cite{dawson2022safe, dai2022learning} synthesize continuous \glspl{cbf} using \glspl{nn} as a function template, which are referred to as \glspl{ncbf}. Yet, these works rely on an initial guess of the forward invariant set or the function structure of the \gls{cbf} to synthesize the \gls{ncbf}. An improper initial guess usually results in a suboptimal \gls{ncbf}. Constructing an optimal \gls{cbf} that renders a maximum forward invariant set is challenging. A recent work \cite{choi2021robust} introduced the \gls{cbvf} which is a safe value function and renders the maximum forward invariant set for a chosen time span. In this work, we synthesize a continuous \gls{ncbf} that approximates the infinite-horizon \gls{cbvf} and renders a safe set that is close to the maximum forward invariant set. 

Although utilizing \glspl{nn} as \glspl{cbf} offers universal approximation capabilities, it necessitates their certification as \glspl{cbf} to provide safety guarantees.
Verifying the \gls{nn} as an \gls{ncbf} in the continuous state space presents a significant challenge. Specifically, since the \gls{nn} is trained using a finite set of data points, it will only be verified on those points. Outside the certified points, safety is no longer guaranteed. There are works \cite{peruffo2021automated, NEURIPS2019_2647c1db} that use the \gls{smt} to verify their \glspl{nn}. However, they are restricted to very simple \glspl{nn} due to expensive computation. 
In this work, we leverage bound propagation techniques~\cite{zhang2018efficient} and the \gls{bbs} to efficiently verify \glspl{ncbf}. 
In particular, we partition the state space and utilize linear bound propagation techniques to provide lower and upper bounds of the \gls{nn} and its Jacobian. These bounds are used to verify if the \gls{nn} satisfies the conditions to be a \gls{cbf}. The \gls{bbs} is applied to refine the partition to improve scalability and achieve less conservative bounds.
We refer to the above verification scheme as \gls{bbv}. This approach is similar to \cite{mathiesen2022safety}, however, we verify \glspl{cbf} instead of barrier functions. 
To accelerate training, we embed the \gls{bbv} into the training loop to synthesize and verify an \gls{ncbf} simultaneously, which we refer to as \gls{bbvt}, see Fig. \ref{fig: system diagram}. We show the efficiency of our method and the practical use of \glspl{ncbf} on an inverted pendulum and a 2D navigation task in a simulation environment.

\section{Related Work}

Many works use \glspl{cbf} to ensure the safety of a system \cite{nguyen20163d, xu2017realizing, ames2016control}. However, it is non-trivial to construct \glspl{cbf}. In recent years, new techniques emerged to automatically synthesize \glspl{cbf}. For a system with polynomial dynamics, a \gls{cbf} can be obtained by solving a \gls{sos} optimization problem ~\cite{2015arXiv150406002A}. Unfortunately, \gls{sos} scales poorly to higher dimensional systems \cite{srinivasan2021extent}.
To address this shortcoming, \glspl{nn} have been employed to approximate \glspl{cbf}. They are trained by supervised learning \cite{dawson2022safe} \cite{dai2022learning} \cite{srinivasan2020synthesis} or \gls{rl} with the Actor-Critic framework \cite{du2023reinforcement} \cite{yang2023model}. However, the quality of the \gls{ncbf} in those works depends on an initial guess of the forward invariant set, \gls{cbf} candidate, or exploration strategy. An improper initial guess results in 
a conservative \gls{ncbf} with a small forward invariant set. To address the conservativity, in this work, we learn a continuous \gls{ncbf} that renders a safe set close to the maximum forward invariant set. Furthermore, the training does not require an initial guess.

Commonly, \glspl{nn} are trained through backpropagation of the empirical loss on a finite set of data points. 
Therefore, it is important to note that even an empirical loss of zero does not guarantee that the certificate is valid everywhere in the state space. Only a few works have verified their \glspl{nn}, such as \cite{boffi2021learning, NEURIPS2019_2647c1db, peruffo2021automated}, which leverage \gls{smt} to provide \glspl{ce} and guarantee the correctness of the synthesis procedure. However, \gls{smt} is limited to simple \glspl{nn} with around 20 neurons in one or two hidden layers due to the need for expensive computation. In contrast to using \gls{smt} for exact verification, several efficient \gls{nn} verification methods using linear bound propagation techniques have been developed \cite{weng2018towards, zhang2018efficient}. These bounding methods provide a new direction to verify neural certificates. The work in \cite{mathiesen2022safety} partitions the state space with a \gls{bbs} and verifies the property of the discrete-time stochastic barrier function for each partition leveraging the method in \cite{zhang2018efficient}. Our work extends the \gls{bbs} of \cite{mathiesen2022safety} to \glspl{cbf} for continuous-time deterministic control-affine systems where the control input constraints must be considered and uses the \gls{bbv} scheme to verify the learned continuous \gls{ncbf}.

\section{Problem Formulation}

Given the following continuous-time control-affine system

\begin{equation}
 \dot{x} = f(x) + g(x)u, \quad x(0) = x_0,
\label{eq: dynamic system}
\end{equation}
where $x\in \mathbb{X} \subset \mathbb{R}^n$, $u \in \mathbb{U} \subset \mathbb{R}^m$, $f:\mathbb{R}^n \rightarrow \mathbb{R}^n$ denotes the autonomous dynamics, and $g:\mathbb{R}^n \rightarrow \mathbb{R}^{n\times m}$ denotes the input dynamics. We assume that $f$, $g$ are Lipschitz continuous and $\mathbb{X}, \mathbb{U}$ are compact sets. 

The safety requirement for the system in \eqref{eq: dynamic system} is encoded via a state admissible set $\mathbb{X}_a \subseteq \mathbb{X}$ and a convex input admissible set $\mathbb{U}_a \subseteq \mathbb{U}$. 
A safe system stays in the state admissible set for all time.
To formally define safety, we use $x_{\pi}(t;x_0)$ to refer to a trajectory of the system in \eqref{eq: dynamic system} at time $t$ with initial condition $x_0$ and control policy $u = \pi(x)$. Safety is then defined as:

\begin{definition}[Safety]
    The system in \eqref{eq: dynamic system} is \emph{safe} if $x_{\pi}(t;x_0) \in \mathbb{X}_a$ and $u=\pi(x_{\pi}(t;x_0)) \in \mathbb{U}_a, ~ \forall t \in [ 0, \infty ] $. 
    \label{definition: safety}
\end{definition}

However, it should be noted that $\mathbb{X}_a$ is not safe everywhere as there may not exist a control input that transitions a state close to the boundary towards the interior of $\mathbb{X}_a$. A safe set should have the property that if the system starts in the safe set, it stays inside for all time. Towards formally defining this property,
let a set $\mathcal{C}$ be defined as the \textit{0-superlevel set} of a continuously differentiable function $h: \mathbb{R}^n \rightarrow \mathbb{R}$, \textit{i.e.,}

\[
\mathcal{C} = \{ x \in \mathbb{X}: h(x) \ge 0 \},
\]
\[ 
\partial \mathcal{C} = \{ x \in \mathbb{X}: h(x) = 0 \}.
\]
Then forward invariance and a safe set are defined as follows.

\begin{definition}[Foward invariance]
    The set $\mathcal{C}$ is \emph{forward invariant} if for every $x_0 \in \mathcal{C}$, there exists a control policy $u = \pi(x) \in \mathbb{U}_a $ such that the trajectory of system in \eqref{eq: dynamic system} $x_{\pi}(t;x_0) \in \mathcal{C},~\forall t \in [0, \infty] $.
    \label{definition: forward invariance}
\end{definition}

\begin{definition}[Safe set]
  The set $\mathcal{C}$ is a \textit{Safe Set} if $\mathcal{C}$ is \textit{forward invariant} and $\mathcal{C} \subseteq \mathbb{X}_a$. 
  \label{definition: safe set}
\end{definition}

A \gls{cbf} renders a safe set and can be used to derive safe control inputs. Before defining \glspl{cbf}, we must introduce extended class $\mathcal{K}_{\infty}$ functions. An extended class $\mathcal{K}_{\infty}$ function is a mapping $\alpha: \mathbb{R} \rightarrow \mathbb{R}$ that is strictly increasing and for which $\alpha(0)=0$ holds. We define a continuous \gls{cbf} as:

\begin{definition} [\acrlong{cbf}]
   Let $\mathcal{C} \subseteq \mathbb{X}_a$ be the 0-superlevel set of a continuously differentiable function $h: \mathbb{R}^n \rightarrow \mathbb{R}$, then $h$ is a \gls{cbf} in $\mathbb{X}_a$ for system in \eqref{eq: dynamic system} if there exists an extended class $\mathcal{K}_{\infty}$ function $\alpha$ such that
    \begin{equation}
        \sup_{u \in \mathbb{U}_a} [L_fh(x) + L_gh(x)u ]  \ge -\alpha(h(x)) \label{eq: CBC}
    \end{equation}
    for all $x \in \mathbb{X}_a$, where $L_f$, $L_g$ represent Lie derivatives.  

    \label{definition: control barrier function}
\end{definition}

With the definition of a \gls{cbf}, we may derive sufficient conditions for a safe system. According to the main result in \cite{CBFtheory}, the following theorem holds:

\begin{theorem}[{\cite[Theorem 2]{CBFtheory}}]
     If function $h$ is a \gls{cbf} for the system in \eqref{eq: dynamic system} and $\frac{\partial h}{\partial x} (x) \neq 0 $ for all $x \in \partial \mathcal{C}$, then any Lipschitz continuous controller $\pi(x) \in K_{cbf}(x)$ with 
    \begin{equation}
        K_{cbf}(x) = \{ u \in \mathbb{U}_a : L_fh(x)+L_gh(x)u + \alpha(h(x)) \ge 0 \}.
    \end{equation}
    renders the set $\mathcal{C}$ safe. Additionally, the set $\mathcal{C}$ is asymptotically stable in $\mathbb{X}_a$.

    \label{theorem: control barrier function render c safe}
\end{theorem}

With Theorem \ref{theorem: control barrier function render c safe}, we are able to ensure the safety of the system in \eqref{eq: dynamic system} as long as a \gls{cbf} is found and its gradient does not vanish on $\partial \mathcal{C}$. 

The objective of this work is to automatically synthesize a \gls{ncbf} and verify it for the continuous state space. The problem is defined as follows.
\begin{problem}
 Given the system in \eqref{eq: dynamic system}, state admissible set $\mathbb{X}_a$, convex input admissible set $\mathbb{U}_a$ and  $\alpha(x) = \gamma x $ where $\gamma$ is a positive constant, synthesize a \gls{ncbf} that is denoted by $\hat{h}_w(x)$, where $w$ are the parameters of the \gls{nn}, and renders set $\mathcal{C}$ safe for the system in \eqref{eq: dynamic system}. This is equivalent to
 \begin{subequations}
\begin{align}
    & \hat{\mathcal{C}} \subseteq \mathbb{X}_a \label{eq: condition 2},  \\
    & \text{inequality } \eqref{eq: CBC} \text{ holds in } \mathbb{X}_a, \label{eq: condition 3} 
\end{align}
\end{subequations}
where $\hat{\mathcal{C}}=\{x\in \mathbb{X}: \hat{h}_w(x) \ge 0  \}$ is the 0-superlevel set of the \gls{ncbf}. 
\end{problem}

\begin{remark}
    The condition $\frac{\partial h}{\partial x }(x) \neq 0 $ for all $x \in \partial C$ is omitted since it generally holds in our setting as we only consider a Tanh-based \gls{fcnn}. More specifically, since $\frac{\partial tanh}{\partial x}(x) \in (0, 1]$ for all $x$, the condition is only violated if either $w = 0$ or catastrophic cancellation occurs in the linear layers, which will almost surely never happen.
\end{remark}

\section{Neural Control Barrier Function Training and Verification}
\label{sec: Verified Neural Control Barrier Function}

In this work, we design a new empirical loss to synthesize an \gls{ncbf}, which is introduced in Section \ref{sec: learning a nCBF}. As the training set only contains a finite set of data points, the \gls{cbf} conditions may not hold in the continuous state space. Therefore, in Section \ref{sec: verify the learned nCBF}, we present the \gls{bbv} to verify \glspl{ncbf}. 
Nevertheless, it is often necessary to iterate through multiple training and verification cycles before successfully learning an \gls{ncbf}. Thus, we introduce \gls{bbvt} in Section \ref{sec: verification-in-loop training}, which embeds \gls{bbv} in the training loop to accelerate training for certifiability.

\subsection{Learning a Neural Control Barrier Function}
\label{sec: learning a nCBF}

The primary goal of this work is to train an \gls{nn}~$\hat{h}_w(x)$ until it satisfies conditions \eqref{eq: condition 2} and \eqref{eq: condition 3} and render a large forward invariant set.
Towards this end, we leverage the main result in \cite[Theorem 3]{choi2021robust}, where a \gls{cbvf} is shown to recover the maximum safe set subject to safety constraints. Contrary to \cite{choi2021robust}, we are interested in infinite-horizon properties. Thus we extend the time-dependent \gls{cbvf-vi} to the infinite-horizon. Let $h(x)$ denote the infinite-horizon \gls{cbvf} and $\rho(x) : \mathbb{X} \rightarrow \mathbb{R}$ denote the signed-distance function for the set $\mathbb{X}_a$, which is defined as $\rho(x) = \inf_{y \in \mathbb{X}/\mathbb{X}_a} \lVert y - x \rVert $ if $x \in \mathbb{X}_a$ and $\rho(x) = - \inf_{y \in \mathbb{X}_a} \lVert y - x \rVert $ if $x \in \mathbb{X}/\mathbb{X}_a$. The infinite-horizon \gls{cbvf-vi} is defined as  

\begin{equation}
\begin{aligned}
    & 0 =  \min \{ \rho(x) -  h(x), \\ &  \max_{u \in \mathbb{U}_a}  L_fh(x) + L_gh(x)u + \gamma h(x) \}.
    \label{CBVF: infinite CBVF-IV}
\end{aligned} 
\end{equation}

We use an \gls{nn} $\hat{h}_w(x)$ to approximate the infinite-horizon \gls{cbvf} $h(x)$. Then, the empirical loss is defined as follows:

\begin{subequations}
\begin{align}
    \mathcal{L} = &  \frac{1}{N_1} \sum_{x \in \mathbb{X}_{a}} \lVert  \min\{ \rho(x) - \hat{h}_w(x),   \nonumber  \\ 
     &  \sup_{u \in \mathbb{U}_a} L_f\hat{h}_w(x)+L_g\hat{h}_w(x)u + \gamma \hat{h}_w(x) - \lambda \} \rVert \label{CBVF: CBVF-IV loss} \\ 
& + \frac{1}{N_2} \sum_{x \in \mathbb{X} / \mathbb{X}_{a}} \max\{\hat{h}_w(x) + \lambda, 0\}. \label{CBVF: inadmissible loss}  \\ \nonumber 
\end{align}
\label{CBVF: empirical loss}
\end{subequations}
where $\lambda$ is a small positive constant to encourage the strict satisfaction of the conditions. The loss term \eqref{CBVF: CBVF-IV loss} shapes the \gls{nn} to be the solution of the infinite-horizon \gls{cbvf-vi} introduced in \eqref{CBVF: infinite CBVF-IV}, which encourages the satisfaction of condition \eqref{eq: condition 3}. The loss term \eqref{CBVF: inadmissible loss} ensures that the \gls{ncbf} is negative in the inadmissible area $\mathbb{X}/\mathbb{X}_a$, which is equivalent to condition \eqref{eq: condition 2}.  Since the system is control-affine, the optimal solution $u^*$ for $\sup_{u \in \mathbb{U}_a} [L_f\hat{h}_w(x)+L_g\hat{h}_w(x)u] $ must be one of the vertices of $\mathbb{U}_a$. Let $\mathbb{U}_a^{\mathcal{V}}$ denote the vertices of the input admissible set, we choose control input $u^\star = \arg \max_{u \in\mathbb{U}_a^{\mathcal{V}}} L_g\hat{h}_w(x)u$. However, the Lie derivative of $\hat{h}_w(x)$ in the early training stage may not align with the Lie derivative of the true \gls{cbvf} $h(x)$. This results in an undesirable optimization path and the \gls{nn} can get stuck at deadlock. The occurrence of a deadlock situation signifies that improvements at certain data points cause constraint violations at other data points, as noted in \cite{liu2023safe}. 
To facilitate the training process and avoid deadlocks, we borrow ideas from \cite{NEURIPS2019_2647c1db, dawson2022safe}, which use a nominal controller to guide the training. Here, we train another neural network $\hat{h}_{\phi}$ with the same structure as $\hat{h}_w$ based on the loss of \cite{fisac2019bridging} and choose $u^* = \arg \max_{u \in \mathbb{U}_a^{\mathcal{V}}} \hat{h}_{\phi}(x + (f(x) + g(x)u)\Delta t)$ by simulating one step ahead to guide the training of the \gls{ncbf} $\hat{h}_w$.

\subsection{Verifying the learned Neural Control Barrier Function}
\label{sec: verify the learned nCBF}

Since the \gls{nn} is trained on finite data points, one must note that the \gls{nn} may not satisfy the \gls{cbf} conditions everywhere in the state space, even if the empirical loss decreases to zero. In fact, condition \eqref{eq: condition 3} may be violated almost everywhere, which means the \gls{nn} may fail to render a forward invariant set and the safety guarantee no longer exists. In this section, we propose to use the \gls{bbv} to verify the learned \gls{ncbf} in the continuous state space.
Specifically, our primary goal is to verify the satisfaction of conditions \eqref{eq: condition 2} and \eqref{eq: condition 3}. 

\begin{figure}
    \centering
    \includegraphics[width=0.85\linewidth]{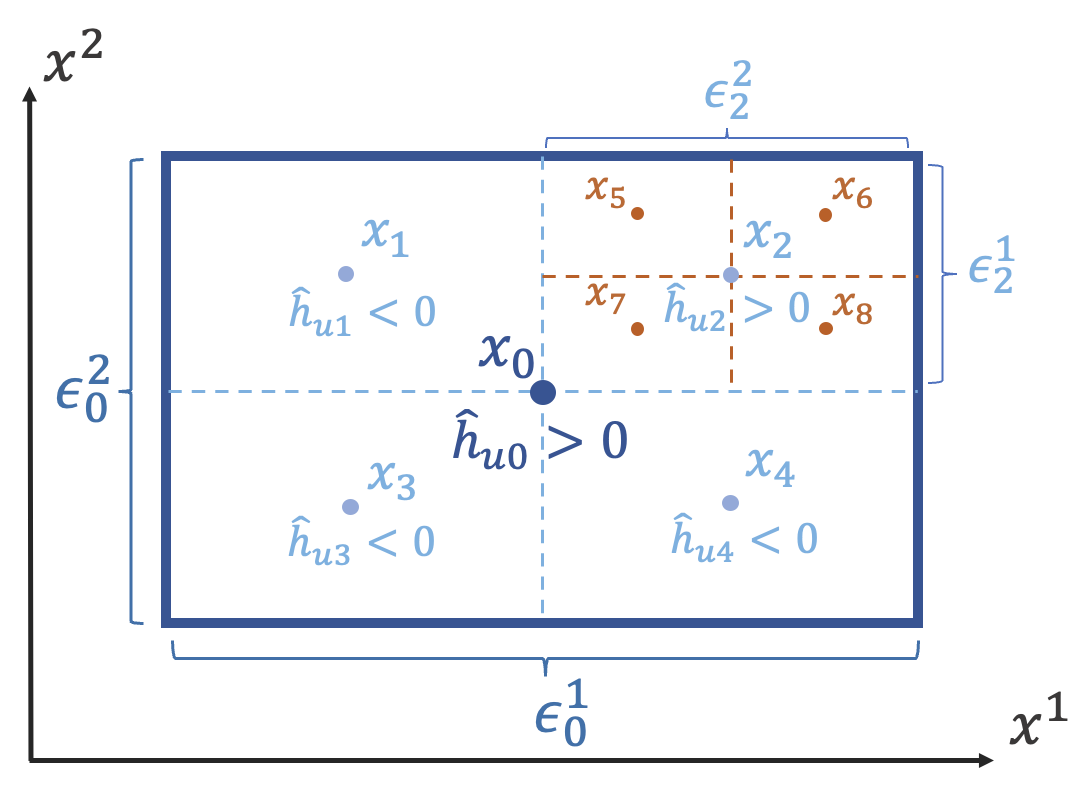}
    \caption{
    An example of the \gls{bbv} in a 2D state space.
    The scheme starts with a coarse partition $\mathbb{B}(x_0, \epsilon_0)$ and refines it using the \acrlong{bbs}. For each hyperrectangle $\mathbb{B}(x_i, \epsilon_i)$, $i = 0, 1, 2, \ldots$, upper bounds for the neural network are computed. In this case, the hyperrectangles  $\mathbb{B}(x_0, \epsilon_0)$ and $\mathbb{B}(x_2, \epsilon_2)$ are refined as $\hat{h}_{u0} > 0$,  $\hat{h}_{u2} > 0$.
    \label{fig: partition state space} }
\end{figure}

Before we explain our verification scheme in detail, we introduce some notations first. Let the partition of the state space be denoted as hyperrectangles $\mathbb{B}(x_i, \epsilon_i) = \{x : \ \mid x -x_i \mid \le \epsilon_i \}$ centered at point $x_i \in \mathbb{X} $ with radius $\epsilon_i \in \mathbb{R}^n$, see Fig. \ref{fig: partition state space}. 
Initially, all hyperrectangles have the same radius $\epsilon_i = \epsilon_{init}$. Let $\mathcal{B} = \{  \mathbb{B}(x_0, \epsilon_0), \dots, \mathbb{B}(x_N, \epsilon_N)\}$ denote the set of all hyperrectangles, $\mathcal{B}_{\mathbb{X}/\mathbb{X}_a} \subset \mathcal{B}$ denote the set of hyperrectangles that covers the inadmissible area $\mathbb{X}/\mathbb{X}_a$, and $\mathcal{B}_{\mathbb{X}_a} \subset \mathcal{B}$ denote the set of hyperrectangles that covers the admissible area.

To verify condition \eqref{eq: condition 2}, which is equivalent to $\hat{h}_w(x) < 0, \forall x \in \mathbb{X}/\mathbb{X}_a$, we rely on the linear bounds of the \gls{nn} computed using CROWN \cite{zhang2018efficient}. The linear bounds are defined as follows:

\begin{equation}
   \hat{h}_{li} \le \hat{h}_w(x) \le  \hat{h}_{ui}, \, x\in \mathbb{B}(x_i,  \epsilon_i).
\end{equation}

We use these linear bounds to certify the satisfaction of condition \eqref{eq: condition 2}. In particular, the upper bound $\hat{h}_{ui}$ can be used to check for non-positivity
\begin{equation}
 \hat{h}_w(x) \le  \hat{h}_{ui} \le 0 , x\in \mathbb{B}(x_i,  \epsilon_i), \mathbb{B}(x_i,  \epsilon_i) \in  \mathcal{B}_{\mathbb{X}/\mathbb{X}_a}.
   \label{eq: constraints for verifying condition 1}
\end{equation}

However, this upper bound tends to be conservative when $\mathbb{B}(x_i, \epsilon_i)$ covers a large area. Therefore, we leverage the \gls{bbs} that starts from the coarse partition and refines each hyperrectangle when $\hat{h}_{ui} > 0$ until $\hat{h}_{ui} \le 0$ or $\epsilon_i \le t_{gap}$ where $t_{gap} > 0 $ is the minimum partition size, see Fig. \ref{fig: partition state space}. If condition \eqref{eq: constraints for verifying condition 1} holds for all hyperrectangles in $\mathcal{B}_{\mathbb{X}/\mathbb{X}_a}$, then the condition \eqref{eq: condition 2} holds in the continuous state space. 

Although verifying condition \eqref{eq: condition 2} is simple, verifying condition \eqref{eq: condition 3} $ \sup_{u\in \mathbb{U}_a} [L_f\hat{h}_w(x)+L_g\hat{h}_w(x)u] \ge - \gamma \hat{h}_w(x), ~\forall x \in \mathbb{X}_a$ is challenging. For improved readability, we denote $q(x) = \sup_{u \in \mathbb{U}_a} [L_f\hat{h}_w(x)+L_g\hat{h}_w(x)u  + \gamma \hat{h}_w(x)]$. 
Hence, verifying condition \eqref{eq: condition 3} is equivalent to verifying $q(x) \ge 0, \forall x \in \mathbb{X}_a$. Let $q_{li}$ define a lower bound of $q(x)$ for $ x \in \mathbb{B}(x_i,  \epsilon_i)$. Then the following condition has to hold:

\begin{equation}
  q(x) \ge q_{li} \ge 0 , ~ x \in \mathbb{B}(x_i,  \epsilon_i), \mathbb{B}(x_i,  \epsilon_i) \in  \mathcal{B}_{\mathbb{X}_a}.
   \label{eq: lower bound of q(x)}
\end{equation}

Similarly to condition \eqref{eq: condition 2}, the \gls{bbs} starts from a coarse partition and refines each hyperrectangle when $q_{li} < 0$ until $q_{li} \ge 0$ or $\epsilon_i \le t_{gap}$. If condition \eqref{eq: lower bound of q(x)} holds for all hyperrectangles in $\mathcal{B}_{\mathbb{X}_a}$, then condition \eqref{eq: condition 3} holds in the continuous state space.

However, the challenge arises in the computation of $q_{li}$. The computation of $q_{li}$ can be reframed as an optimization problem within the hyperrectangle $\mathbb{B}(x_i, \epsilon_i)$

\begin{subequations}
    \begin{align}
        q_{li} = \min_{x} ~ & q(x)  \label{eq: problem 1}\\
          \mathrm{s.t.} ~ & x \in \mathbb{B}(x_i, \epsilon_i).
    \end{align}
\end{subequations}

The term $q(x)$ is a complex function containing nonlinear dynamic functions $f$, 
$g$, a neural network $\hat{h}_w$ as well as its Jacobian, which renders a constrained \gls{nlp} in \eqref{eq: problem 1}. The state-of-the-art \gls{nlp} solver \cite{Luenberger2021Nov} requires gradients of the objective function, which involves computation of the Hessian of the \gls{nn}. The expensive computation makes it impractical to solve \eqref{eq: problem 1} directly. 

Although computing the lower bound of $q(x)$ is quite complex, computing the bound of the components of $q(x)$ separately is much simpler. We can compute the bound of the \gls{nn} using CROWN \cite{zhang2018efficient} and its Jacobian leveraging a recent result in \cite{shi2022efficiently} or \cite{laurel2022dual}:

\begin{gather}
   \hat{h}_{li} \le \hat{h}_w(x)  \le  \hat{h}_{ui} , \forall x\in \mathbb{B}(x_i, \epsilon_i),
   \label{eq: linear constraints on NN}\\
    J_{li} \le \nabla \hat{h}_w(x) \le J_{ui}, \forall x \in \mathbb{B}(x_i, \epsilon_i).
    \label{eq: bound on jacobian}
\end{gather}

Furthermore, we can approximate the nonlinear dynamic functions $f$ and $g$ using Taylor Models as done in \cite{streeter2022automatically} or sampling:

\begin{equation}
    x_{li} \le  f(x) + g(x) u^{\star}  \le x_{ui}, \forall x \in \mathbb{B}(x_i, \epsilon_i).
    \label{eq: linear constraints on dxdt}
\end{equation}

In \eqref{eq: problem 1}, the objective function depends on the variable $x$ and is constrained within the feasible region for $x$.  We simplify \eqref{eq: problem 1} by considering three independent variables subject to independent constraints. This results in

\begin{subequations}
    \begin{align}
        q'_{li}  = \min_{h, J, x} & \  q'(h, J, x) = \langle J, x\rangle + \gamma h
        \label{eq: problem 2} \\
         \mathrm{s.t.} & \  \hat{h}_{li} \le h \le \hat{h}_{ui}, \\
              & \ J_{li} \le  J \le J_{ui}, \\
              & \   x_{li} \le x \le x_{ui},
    \end{align}
\end{subequations}
where $x$ denotes the value of $f(x)+g(x)u$, $h$ denotes the value of $\hat{h}_w(x)$ and $J$ denotes the value of $\nabla \hat{h}_w(x)$. 
When \eqref{eq: linear constraints on NN}, \eqref{eq: bound on jacobian}, and \eqref{eq: linear constraints on dxdt} are over-approximations of the true intervals, it is clear that the optimal solution $q'_{li}$ from \eqref{eq: problem 2} is an over-approximation of the optimal solution $q_{li}$ from \eqref{eq: problem 1},  which means $q'_{li} \le q_{li}$. 
To efficiently solve \eqref{eq: problem 2}, we may compute the optimal solution independently for each term, taking the minimum over the set of vertices.

Although the theoretical complexity of the \gls{bbv} is still exponential in the dimension of the state space, it improves the scalability in practice. One must note that our method is a sound verification method instead of a complete one, which means the failure to obtain $\mathcal{B}_{\mathbb{X}/\mathbb{X}_a}$ and $\mathcal{B}_{\mathbb{X}_a}$ that satisfy condition \eqref{eq: constraints for verifying condition 1}, \eqref{eq: lower bound of q(x)} does not mean the invalidation of the \gls{ncbf}, as we over-approximate the conditions.
Note that although the universal approximation theorem in \cite{hornik1989multilayer} guarantees the existence of $\hat{h}_w(x)$ to be an \gls{ncbf} that renders maximum forward invariant set, this is under the assumption that the \gls{nn} has a sufficient number of neurons.

\subsection{Branch and Bound Verification-in-the-loop Training}
\label{sec: verification-in-loop training}

Although the \gls{bbv} provides a practical way to certify the \gls{nn} as \gls{ncbf}, it requires several training and verification processes until an \gls{ncbf} is obtained. Therefore, leveraging the information from the verification and ensuring the satisfaction of conditions \eqref{eq: condition 2} and \eqref{eq: condition 3} becomes the task of \gls{bbvt}. This type of method is also known as \gls{ce} guided inductive synthesis \cite{abate2018counterexample}. See Fig.~\ref{fig: system diagram} for an overview of the framework.

We start with the initial fixed training dataset $\mathcal{D}$ that contains a number of uniformly sampled points. During the training procedure, we optimize the \gls{nn} to decrease the loss in \eqref{CBVF: empirical loss} using $\mathcal{D}$. After $k$ epochs, the verifier starts with a coarse partition of the state space. The upper bound $\hat{h}_{ui}, ~\forall \, \mathbb{B}(x_i, \epsilon_i) \in \mathcal{B}_{\mathbb{X}/\mathbb{X}_a}$ and lower bound $ q'_{li}, ~\forall \, \mathbb{B}(x_i, \epsilon_i) \in \mathcal{B}_{\mathbb{X}_a}$ are computed. The hyperrectangles, whose $\hat{h}_{ui} \ge 0 $ or $ q'_{li}\le 0$, are split until $\epsilon_i \le t_{gap}$. After reaching the minimum partition size $t_{gap}$, the hyperrectangles whose $\hat{h}_{ui} \ge 0 $ or $ q'_{li} \le 0$ are treated as the violation areas. The center points are added to the \gls{ce} dataset and the training procedure is repeated until the verifier returns \texttt{satisfaction}  or the maximum number of iterations $n_\mathrm{max}$ is reached.

The training procedure is not guaranteed to converge to an \gls{ncbf}, but if the verifier returns \texttt{satisfaction}, the \gls{nn} is an \gls{ncbf} for the given system in the continuous state space. 

\section{Results}

In this section, we evaluate our proposed framework on two systems: an inverted pendulum and a 2D navigation task. The experimental setup is introduced in Section~\ref{sec:ex_setup}. In Section~\ref{sec:verification} and Section~\ref{sec: Optimality}, we provide a comprehensive assessment on the inverted pendulum, addressing the verification efficiency and the size of the safe set, respectively.
In Section~\ref{sec:results_2dnav}, we consider a 2D navigation task with nonconvex constraints to display the practical use of our framework and combine the learned \gls{ncbf} with \gls{rl} to achieve safe learning.

We consider the following baseline methods:
\begin{itemize}
    \item LST:  The Level Set Toolbox (LST)  \cite{mitchell2007toolbox} generates a safe value function by \gls{hji-ra} over a discrete grid. 
    \item NeuralCLBF: Neural Control Lyapunov Barrier Function (NeuralCLBF) \cite{dawson2022safe} parametrizes the \gls{cbf} as an \gls{nn} and optimizes it according to their empirical loss based on \eqref{eq: CBC} and a nominal safe set.
    \item SMT: \cite{NEURIPS2019_2647c1db} trains a neural Lyapunov function with \gls{smt} generating counterexamples and ensures the validation of the result. The constraints considered by \gls{smt} are conditions \eqref{eq: condition 2} and \eqref{eq: condition 3}. To have a fair comparison, the training loss is chosen to be the same as in \eqref{CBVF: empirical loss}.
\end{itemize}

\subsection{Experimental Setup}\label{sec:ex_setup}

\subsubsection{Inverted Pendulum}

\begin{figure}[b]
\vspace{-5pt}
    \centering  \includegraphics[width=0.65\linewidth]{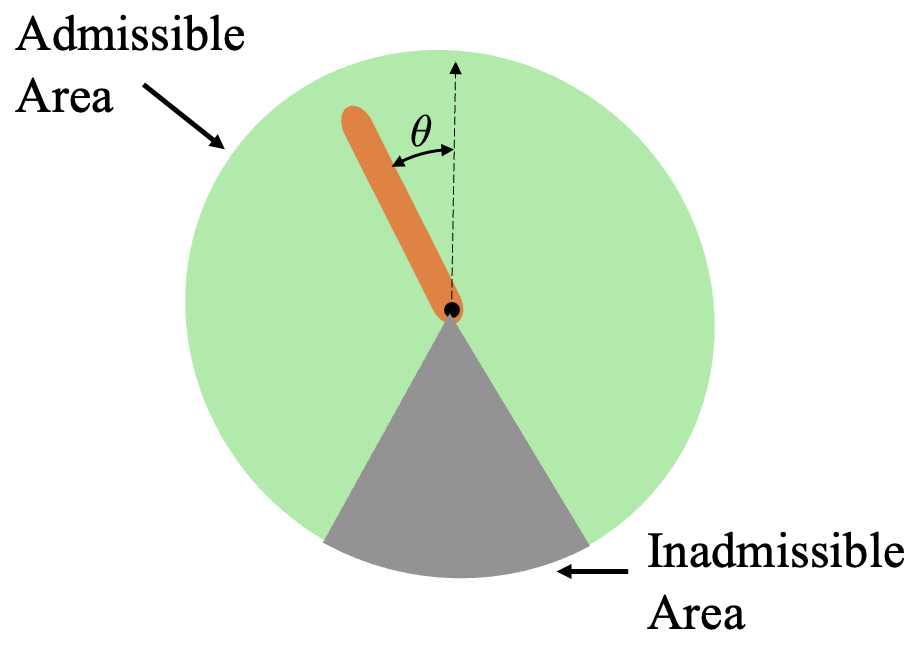}
    \caption{The workspace of the considered inverted pendulum.}
    \label{fig: inverted pendulum workspace}
\end{figure}

Let $s=[\theta, \dot{\theta}] \in \mathbb{X} \subset \mathbb{R}^2 $ be the state variable and $u \in \mathbb{U} \subset \mathbb{R}$ be the control input. We consider the state space $\mathbb{X} =\{ s : [-\pi, -5] \le s \le [\pi, 5]\}$ and the input space $\mathbb{U} = \{ u : -12 \le u \le 12 \}$. The dynamics of the inverted pendulum are given by:

\begin{equation}
    \begin{split}
        \dot{\theta} & = \dot{\theta}, \\
        \ddot{\theta} & = \frac{3g}{2l}sin(\theta) - \frac{3\beta}{ml^2}\dot{\theta} + \frac{3}{ml^2}u,
    \end{split}
    \label{eq: dynamics of inverted pendulum}
\end{equation}
where $m=1$, $b=0.1$, $g= 9.81$, and $l=1$. The state admissible set is  $\mathbb{X}_a=\{s : [-\frac{5\pi}{6}, -4] \le s \le  [\frac{5\pi}{6}, 4] \}$ and the input admissible set is $\mathbb{U}_a=\mathbb{U}$, see Fig. \ref{fig: inverted pendulum workspace}. 

\subsubsection{2D navigation task}\label{sec:2dNav}
We consider a 2D navigation task in which a point robot should reach a goal position while avoiding obstacles, see Fig. \ref{fig: trajectories during training without nCBF}. Let $s = [x, y, \dot{x}, \dot{y}] \in \mathbb{X}  \subset  \mathbb{R}^4$ be the state variable and $u = [a_x, a_y] \in \mathbb{U}  \subset \mathbb{R}^2 $ be the control input representing the acceleration along the x-axis and y-axis. The dynamics of the point robot are:

\begin{equation}
    \begin{bmatrix}
        \dot{x} \\ \dot{y} \\ \ddot{x} \\ \ddot{y}
    \end{bmatrix} = 
    \begin{bmatrix}
        0 & 0 & 1 & 0 \\
        0 & 0 & 0 & 1 \\
        0 & 0 & 0 & 0 \\
        0 & 0 & 0 & 0
    \end{bmatrix}
    s^T + 
    \begin{bmatrix}
        0 & 0 \\
        0 & 0 \\
        1 & 0 \\
        0 & 1 \\
    \end{bmatrix}
    \begin{bmatrix}
        a_x \\
        a_y
    \end{bmatrix}
    \label{eq: dynamics of point robot}
\end{equation}

We consider the obstacle sets $X^1 = \{ s : 0 \le x \le 4 \ \wedge \ 0 \le y \le 4 \}$, $X^2 = \{ s : 1.5 \le x \le 2.5 \ \wedge \ 0 \le y \le 2 \}$, speed constraints $X^3 = \{ s : -1 \le \dot{x} \le 1 \ \wedge \ -1 \le \dot{y} \le 1 \}$, and input constraints $U^1 = \{ u : -1 \le a_x \le 1 \ \wedge \ -1 \le a_y \le 1 \}$. Thus, the state admissible set is $\mathbb{X}_a = X^1 \cup (X^2)^C \cup X^3$, where $(.)^C$ represents the complement of a set, and the input admissible set is $U_a = U^1$.

\subsubsection{Training Configuration}

\begin{table}
\small
\caption{Hyper-parameter for \gls{ncbf} Training.}
\resizebox{\columnwidth}{!}{%
\begin{tabular}{|c|c|c|c|}
\hline
$\gamma$ &
  0.5 &
  $\lambda$ &
  0.05 \\ \hline
learning rate $r$ &
  $10^{-3}$ &
  \begin{tabular}[c]{@{}c@{}}learning rate \\ decay $\beta$\end{tabular} &
  0.995 \\ \hline
\begin{tabular}[c]{@{}c@{}}verify after \\ every $k$ epochs\end{tabular} &
  20 &
  \begin{tabular}[c]{@{}c@{}}minimum partition\\  gap $t_{gap}$\end{tabular} &
  0.005 \\ \hline
\begin{tabular}[c]{@{}c@{}}initial radius $\epsilon_{init}$ \\ (inverted pendulum)\end{tabular} &
  {[}0.2, 0.2{]} &
  \begin{tabular}[c]{@{}c@{}}initial radius $\epsilon_{init}$ \\ (2D navigation)\end{tabular} &
  \begin{tabular}[c]{@{}c@{}}{[}0.2, 0.2, \\ 0.2, 0.2{]}\end{tabular} \\ \hline
\begin{tabular}[c]{@{}c@{}}Num. fixed points\\ (inverted pendulum)\end{tabular} &
  $10^5$ &
  \begin{tabular}[c]{@{}c@{}}Num. fixed points\\ (2D navigation)\end{tabular} &
  $10^6$ \\ \hline
  $n_{max}$ &
  100 &
   &
   \\ \hline
\end{tabular}%
}
\label{tab: hyper-parameters}
\vspace{-18pt}
\end{table}
\begin{figure*}
\centering
\begin{subfigure}{0.32\textwidth}
    \centering
    \includegraphics[width=\linewidth]{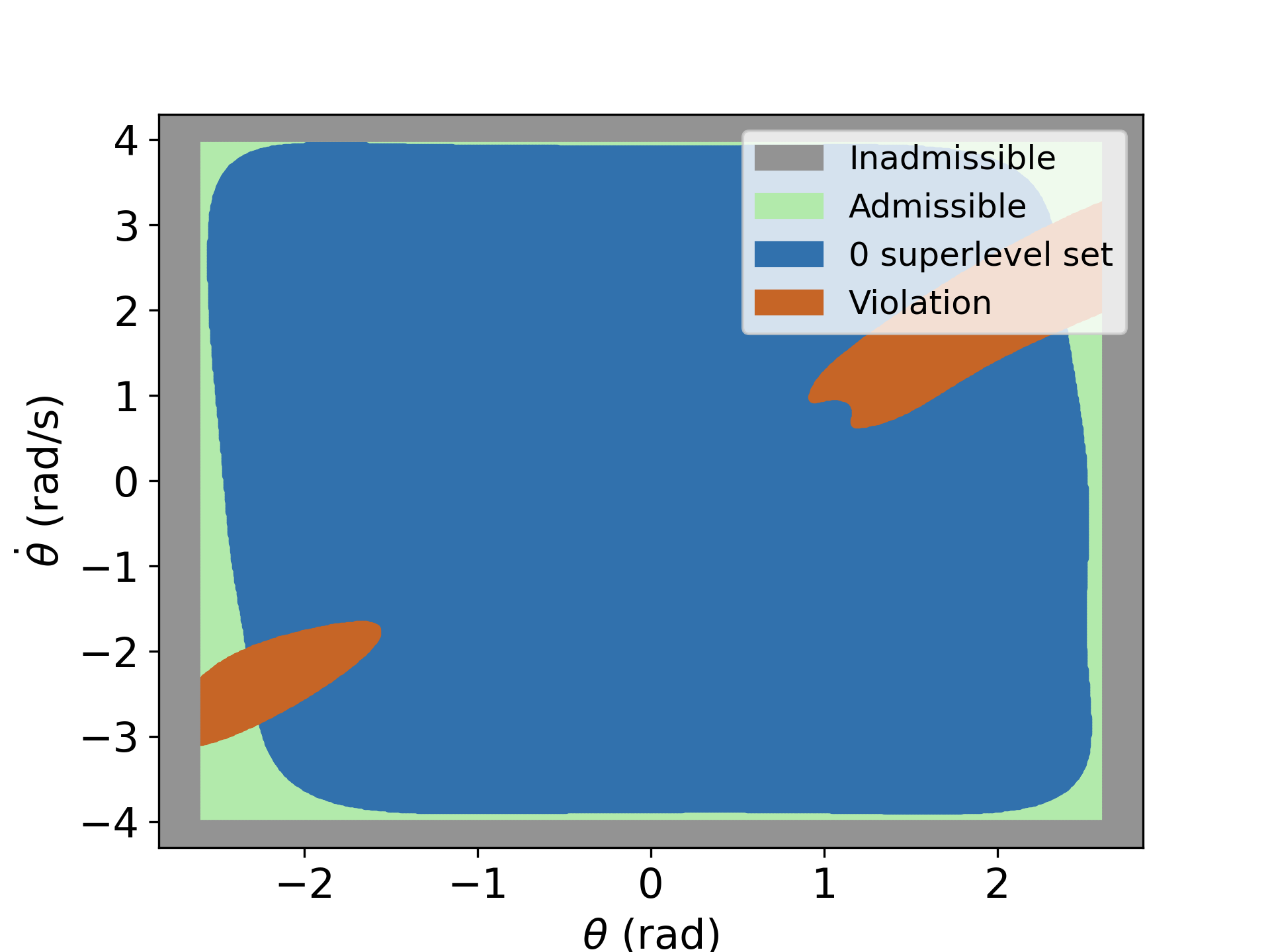}
    \caption{Training without verification-in-the-loop}
    \label{fig: training result without verification-in-loop}
\end{subfigure}
\begin{subfigure}{0.32\textwidth}
    \centering
    \includegraphics[width=\linewidth]{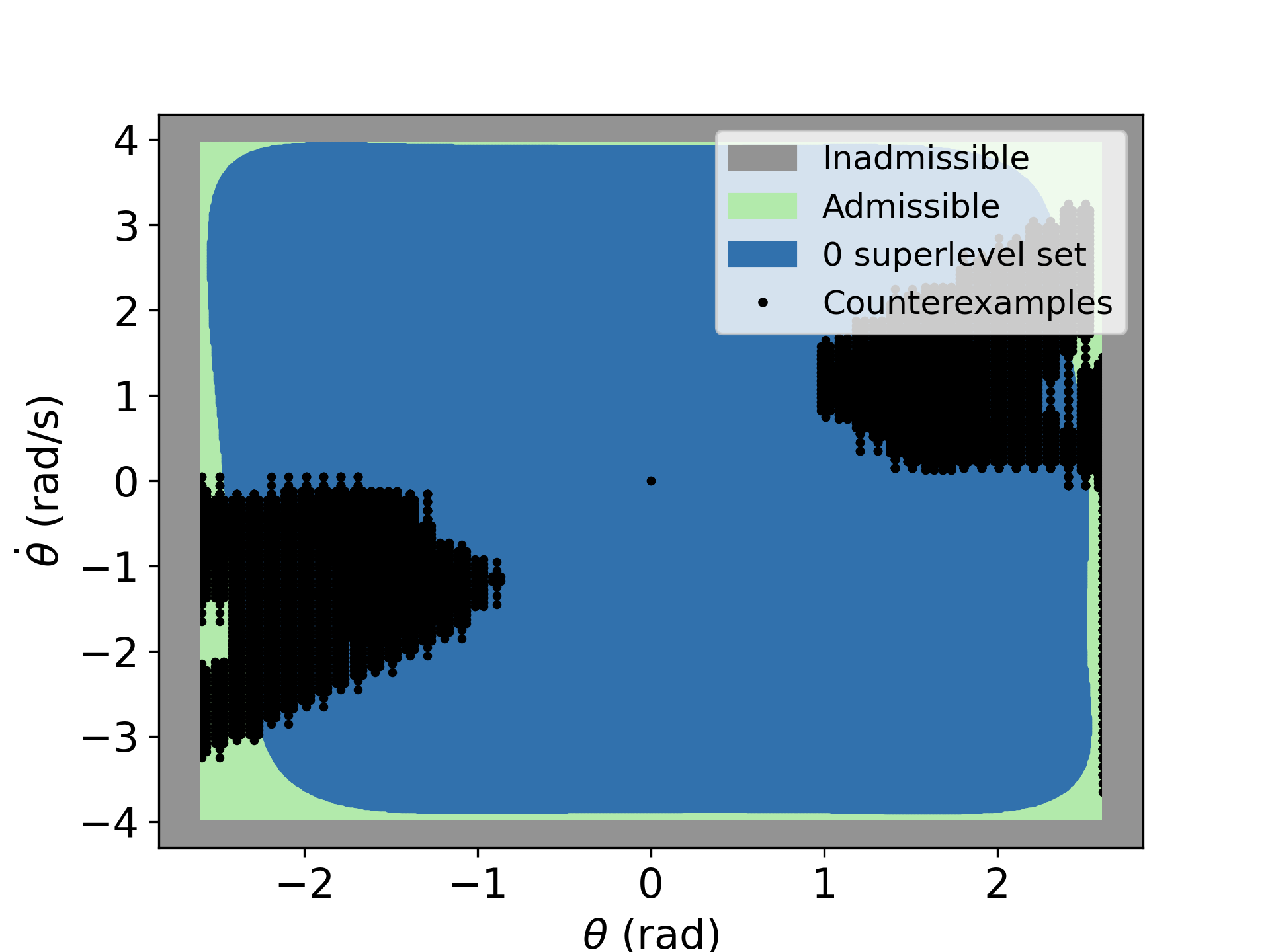}
    \caption{Distribution of counterexamples}
    \label{fig: augmented training data set}
\end{subfigure}
\begin{subfigure}{0.32\textwidth}
    \centering
    \includegraphics[width=\linewidth]{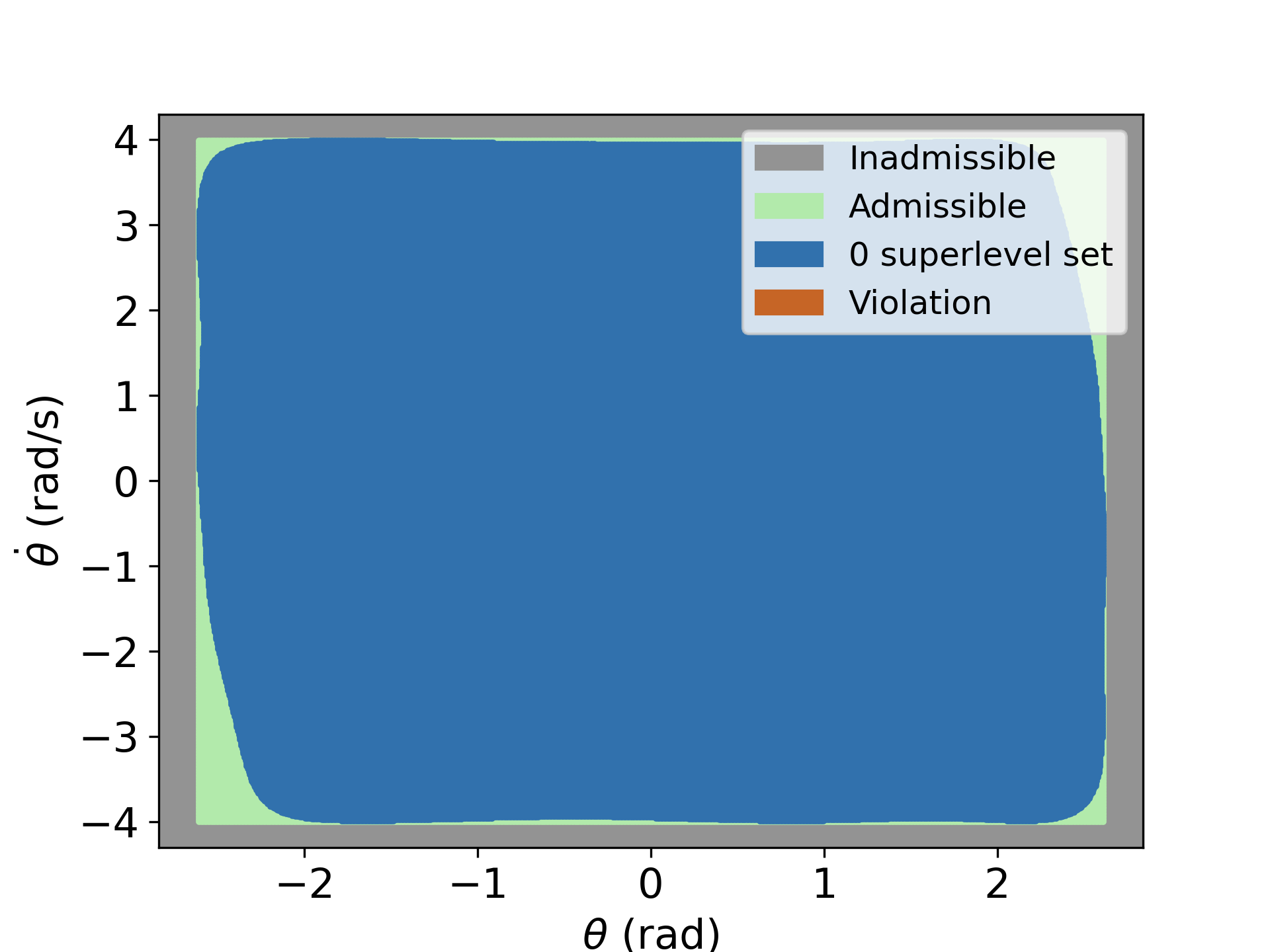}
    \caption{Training with verification-in-the-loop}
    \label{fig: training result with verification-in-loop}
\end{subfigure}
\caption{Shapes of 0-superlevel sets of \glspl{nn} trained with and without \gls{bbvt} for the inverted pendulum. In Fig. \ref{fig: training result without verification-in-loop} the \gls{nn} is trained with a fixed dataset and evaluated on a denser testing dataset to showcase that condition \eqref{eq: condition 3} is not satisfied for the continuous state space. Figure \ref{fig: augmented training data set} shows the counterexamples added to the dataset according to \gls{bbvt}. Figure \ref{fig: training result with verification-in-loop} showcases that, after training the \gls{nn} with \gls{bbvt}, no validations are detected since the \gls{nn} is an \gls{ncbf}.}
\label{fig: verification-in-loop training result}
\vspace{-18pt}
\end{figure*}
For both systems, we train the \gls{ncbf} using Pytorch on NVIDIA A40, and \acrlong{sgd} is used as the optimizer to avoid local minima. The used hyper-parameters can be found in Table~\ref{tab: hyper-parameters}. For the inverted pendulum, we choose a Tanh-based \gls{fcnn} with one hidden layer which consists of 36 neurons. 
For the 2D navigation task, a larger Tanh-based \gls{fcnn} is required since the shape of the environment is more complex. Here we choose a Tanh-based \gls{fcnn} with two hidden layers, each of which consists of 256 neurons.

\subsection{Verification and Efficiency}\label{sec:verification}
In this section, we use the inverted pendulum to discuss the certification of the trained \gls{nn} as an \gls{cbf}. To showcase the disadvantage of training without verification, we train the \gls{ncbf} with a fixed data set and stop training the \gls{ncbf} after $200$ epochs. We then examine the satisfaction of condition~\eqref{eq: condition 3} with a denser testing dataset. The 0-superlevel set of the trained \gls{nn} is shown in blue in Fig. \ref{fig: training result without verification-in-loop}. The orange area indicates the testing data points that violate condition \eqref{eq: condition 3}.

We resume the training with the same dataset and use \gls{bbvt} to augment the training dataset with \glspl{ce} every $k$ epochs until the verifier returns \texttt{satisfaction}. Figure~\ref{fig: augmented training data set} shows the distribution of the \glspl{ce} after the first verification loop. As we augment the dataset, the verifier returns \texttt{satisfaction} after 240 epochs, see Fig.~\ref{fig: training result with verification-in-loop}. 

To highlight the efficacy of \gls{bbvt}, we evaluate the training time, verification time, and the ratio of violation areas for our framework and the baseline methods.
The results are shown in Table~\ref{tab: validation and efficiency comparison}. 
We first compare our method with LST \cite{mitchell2007toolbox}. The table shows the results of LST for two different grid gaps, which are $0.2$ and $0.05$ respectively. It is evident that an increased grid density leads to improved accuracy at the cost of longer computation time. However, a dense grid map is not always possible, since the memory space of LST grows exponentially, which is referred to as the \textit{Curse of Dimensionality}. With a grid gap of $0.05$, LST requires $24.41$kB memory space, while we only need to store the parameters of the \gls{ncbf}, which is $1.2$kB. This is important for embedded devices such as the control unit on drones.

Then, we compare our method with NeuralCLBF. Due to the lack of a verification process and counterexample data set, the fixed data set for NeuralCLBF contains $10^6$ data points in order to have a fair comparison with our method. Since NeuralCLBF learns an \gls{ncbf} based on a nominal safe set, the training process is assisted by prior knowledge and results in less training time, see Table~\ref{tab: validation and efficiency comparison}. However, there are sparse areas that violate the conditions as discussed in \cite{dawson2022safe} and how these sparse areas grow with the complexity of the system has not been studied yet.

We also compare our method with \gls{smt}. However, \gls{smt} did not return \texttt{satisfaction} until the maximum number of iterations $n_\mathrm{max}$ was reached, see Table ~\ref{tab: validation and efficiency comparison}. Although there exist some works \cite{NEURIPS2019_2647c1db, peruffo2021automated} that use \gls{smt} to verify a neural controller, they only use a very simple \gls{fcnn} with around 9 neurons. In our case, the computation time of \gls{smt} grows dramatically since the \gls{nn} is more complex. Also, \gls{smt} can only generate several counterexamples at each iteration, while \gls{bbvt} generates all the counterexamples in state space $\mathbb{X}$, which is more efficient than \gls{smt}.

\begin{table*}
\centering
\caption{Verification and Efficiency Comparison for the inverted pendulum. \gls{bbvt} is compared against LST, NeuralCLBF, and SMT to synthesize an \gls{ncbf}. LST and NeuralCLBF do not verify their safe value function, which is represented by '-' in columns 3 and 4. To validate the verification process, we calculate the ratio of points that violate condition \eqref{eq: condition 3} on a uniform grid with a size of $10^3 \times 10^3$ within the state space $\mathbb{X}$.}
\begin{tabular}{cccccc}
\hline
 &
  Stop criteria &
  \begin{tabular}[c]{@{}c@{}}Total computation time\\ (s)\end{tabular} &
  \begin{tabular}[c]{@{}c@{}}Average verification time\\ (s/epoch)\end{tabular} &
  \begin{tabular}[c]{@{}c@{}}Average generation time\\ (s/per counterexample)\end{tabular}  &\begin{tabular}[c]{@{}c@{}}Violation points/testing points\\ ($\%$)\end{tabular} \\ \hline
LST(0.2)        & value converges & 5.34 & -      & -   &  1.9 \\
LST(0.05)        & value converges & 104.48 & -      & -   & 0.0064 \\
NeuralCLBF        & loss converges & 584.6 & -      & -   & 0.0013  \\
SMT          & max \# iter reached         & 5311.68     & 14.73 & 1.34 & 0.7742 \\ 
BBVT(ours)  & verified      & 1214.15 & 16.20 & 0.004 & 0.0 \\ \hline
\end{tabular}
\label{tab: validation and efficiency comparison}
\vspace{-15pt}
\end{table*}

\subsection{Size of Safe Set}
\label{sec: Optimality}
We will compare the size of the forward invariant set derived using our framework and the baseline methods in this section. Since \gls{smt} failed to verify the \gls{ncbf} and LST with a grid gap of $0.2$ has a large violation area, we compare our framework only against LST with a grid gap of $0.05$ and NeuralCLBF with the nominal safe set being $\mathbb{X}_n = \{s : \lVert s \rVert  < \frac{3 \pi}{4} \}$). 

\begin{figure}
    \centering    \includegraphics[width=0.85\linewidth]{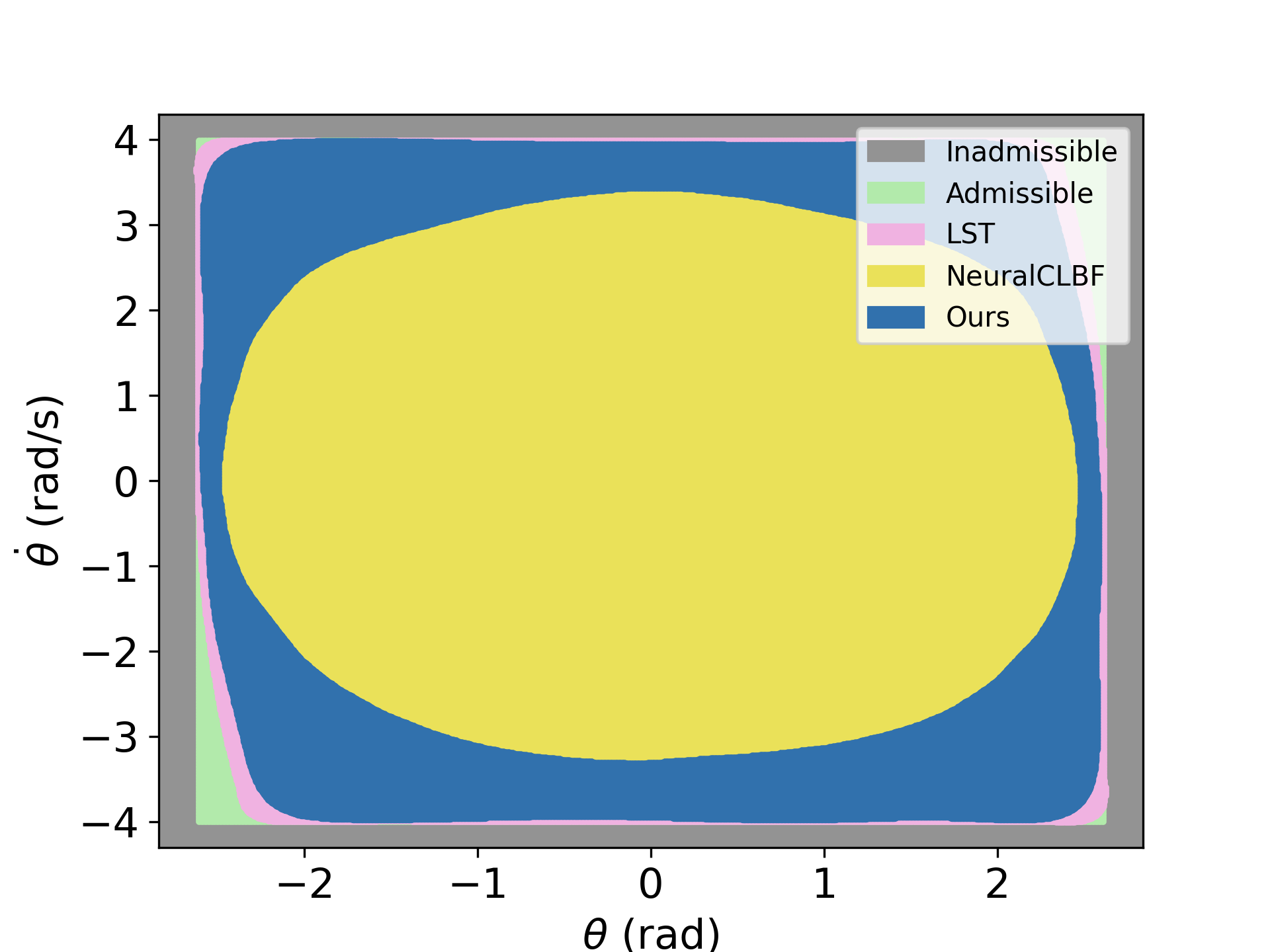}
    \caption{The forward invariant set of safe value functions obtained by different methods for the inverted pendulum.}
     \label{fig: inverted pendulum control invariant set}
\vspace{-20pt}
\end{figure}

The forward invariant sets derived by the different methods are illustrated in Fig. \ref{fig: inverted pendulum control invariant set}. We see that the size of the forward invariant set from NeuralCLBF is conservative, while our method approximates \gls{cbvf} and renders a safe set that is close to the maximum forward invariant set.
Since the forward invariant set of the \gls{cbf} is always a subset of that from \gls{hji-ra}, which is discussed in \cite{choi2021robust}, it is not surprising that LST renders a larger safe set than ours. 
We note that $\lambda > 0$ in \eqref{CBVF: CBVF-IV loss} encourages the satisfaction of the \gls{cbf} conditions at the expense of rendering a smaller safe set. 

\subsection{Application of Neural Control Barrier Functions to Safe Learning}\label{sec:results_2dnav}

\begin{figure*}
\centering
\begin{subfigure}{0.36\textwidth}
    \centering\includegraphics[width=\linewidth]{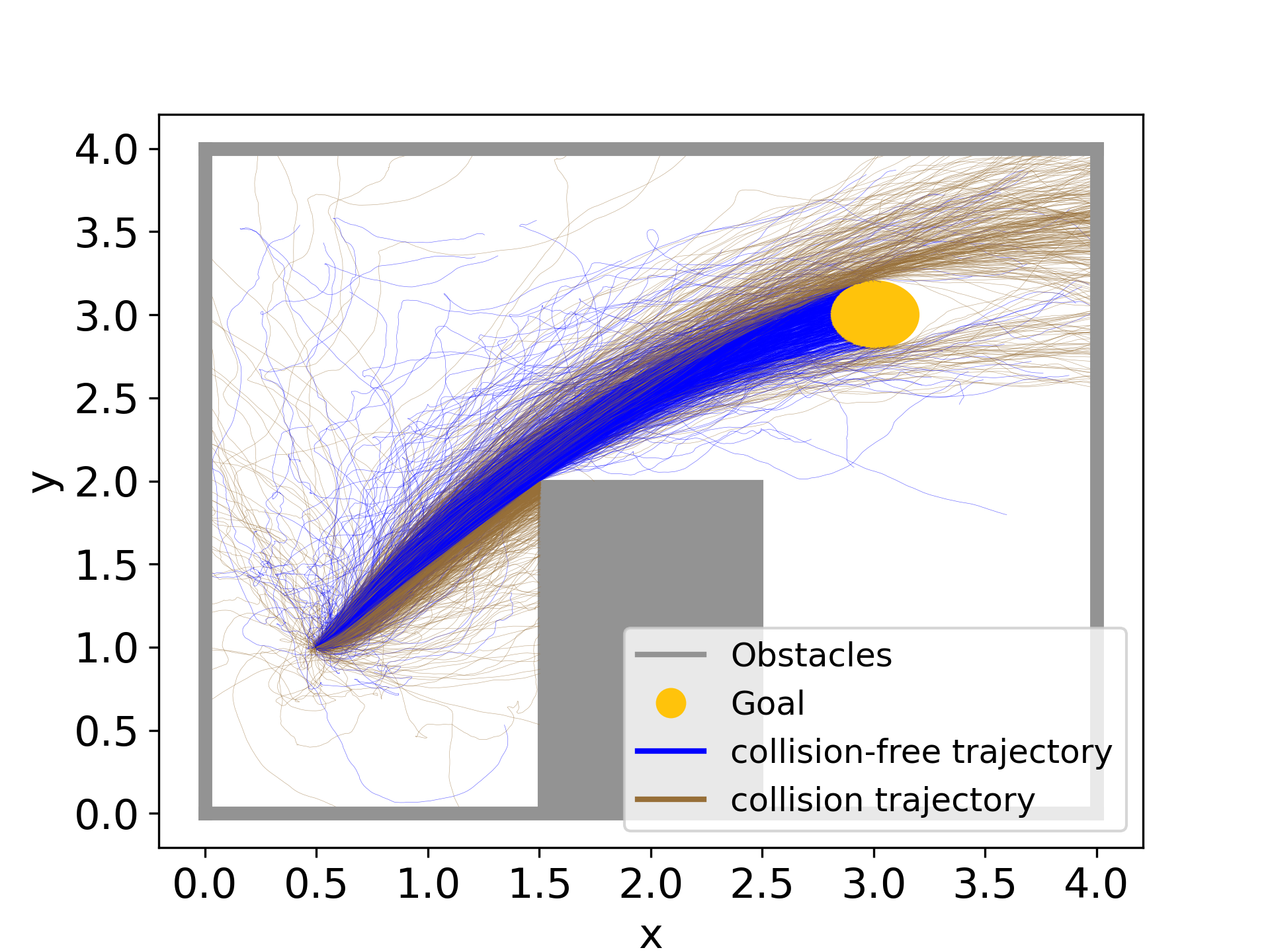}
    \caption{Trajectories during learning without \gls{ncbf}}
    \label{fig: trajectories during training without nCBF}
\end{subfigure}%
\begin{subfigure}{0.36\textwidth}
    \centering
    \includegraphics[width=\linewidth]{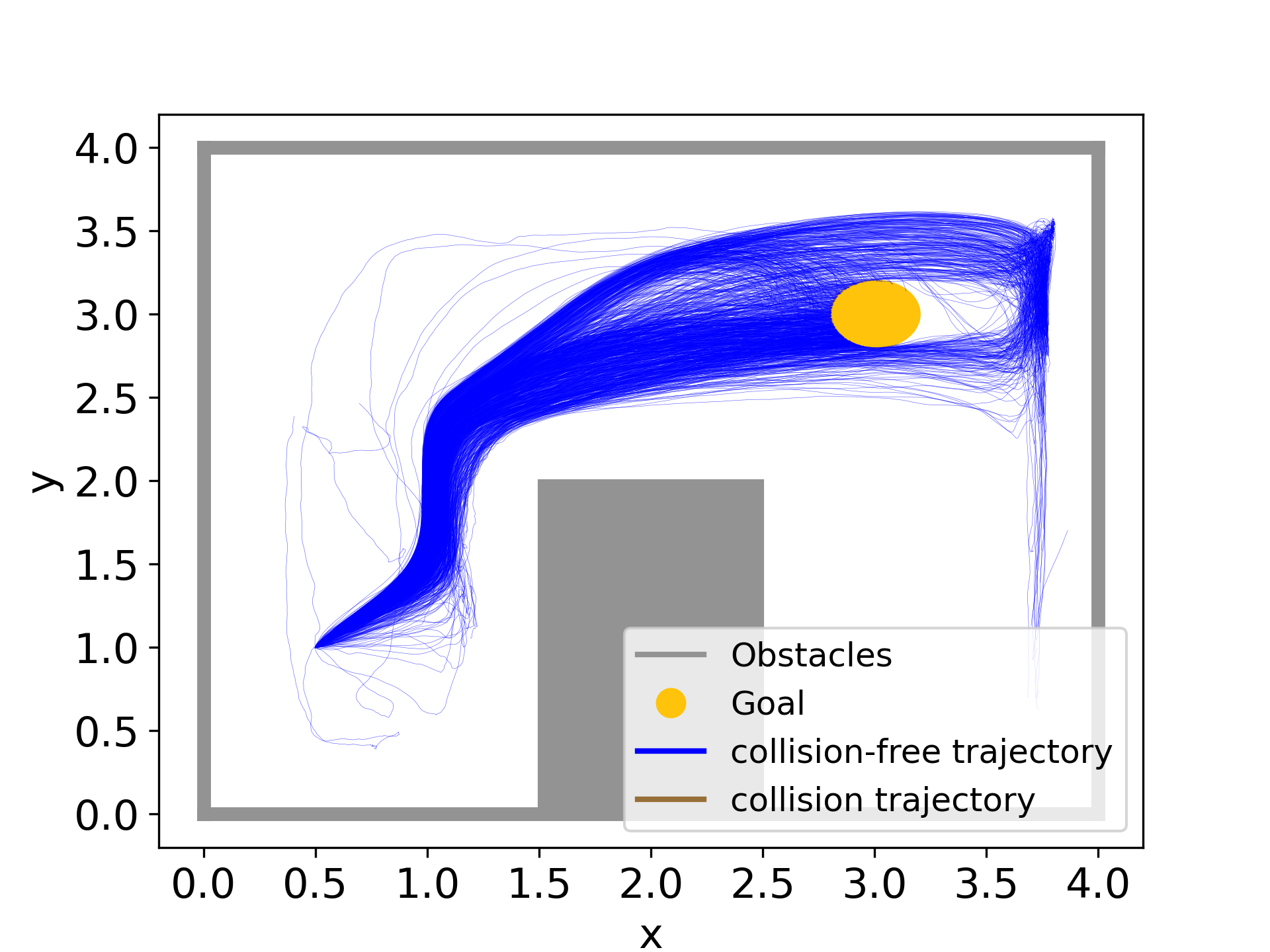}
    \caption{Trajectories during learning with \gls{ncbf}}
    \label{fig: trajectories during training with nCBF}
\end{subfigure}%

\begin{subfigure}{0.36\textwidth}
    \centering
    \includegraphics[width=\linewidth]{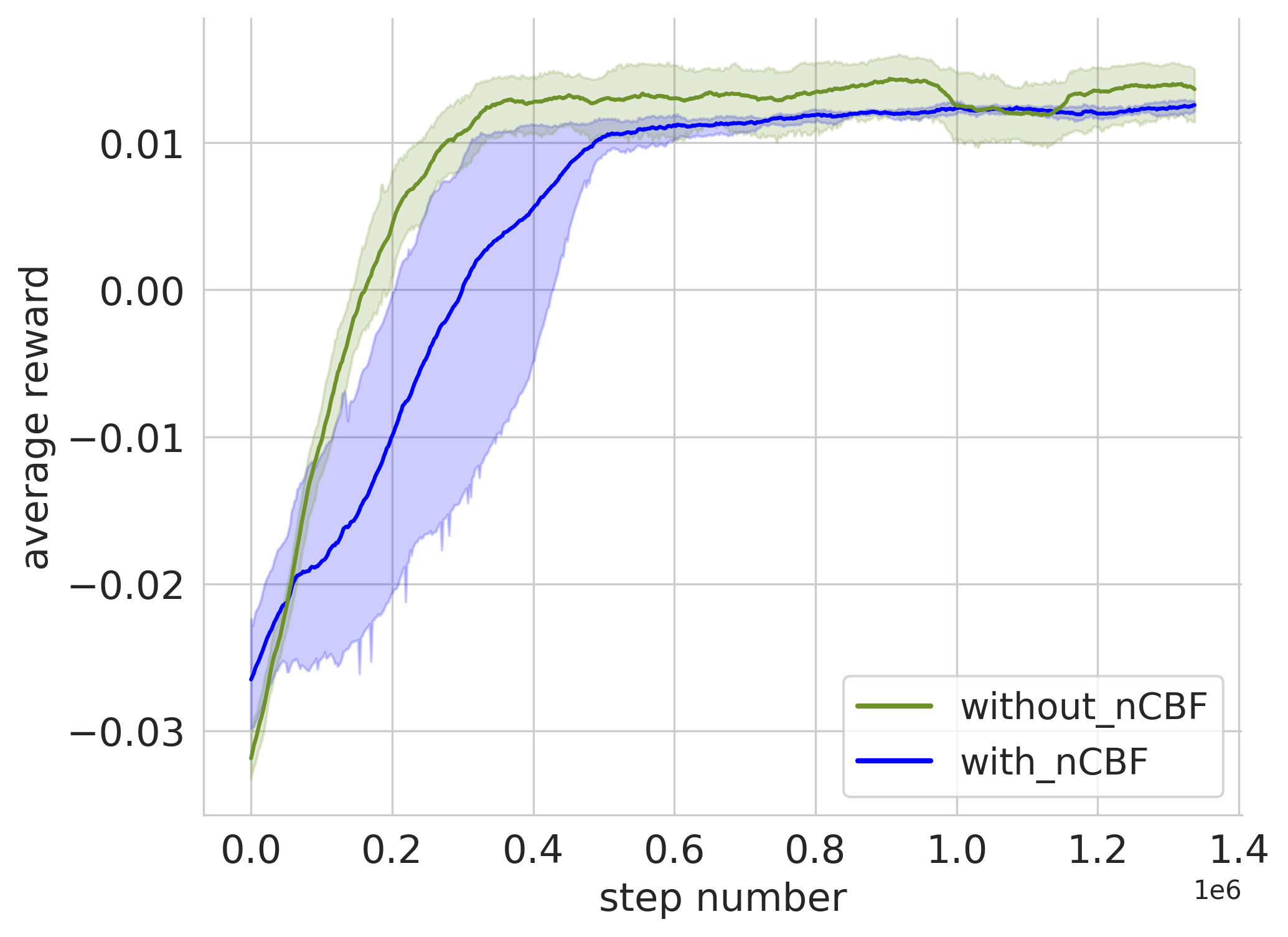}
    \caption{Moving average reward}
    \label{fig: average reward during training}
\end{subfigure}
\begin{subfigure}{0.36\textwidth}
    \centering
    \includegraphics[width=\linewidth]{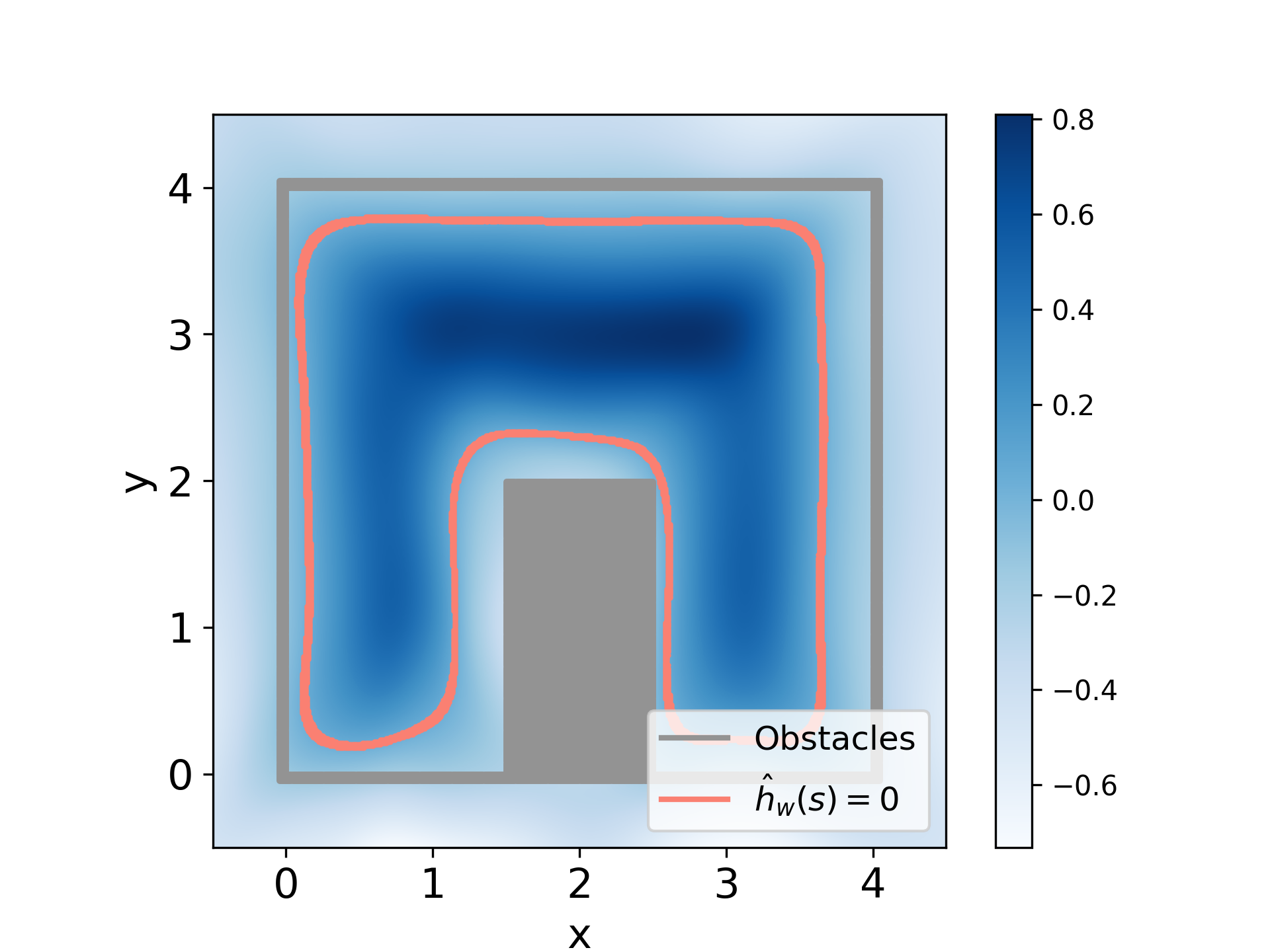}
    \caption{Heatmap of the \gls{ncbf}.}
    \label{fig: contour of nCBF}
\end{subfigure}%
\caption{\ref{fig: trajectories during training without nCBF} and \ref{fig: trajectories during training with nCBF} show all the trajectories during \gls{rl} training. \ref{fig: average reward during training} illustrates the moving average reward of every 2048 steps. \ref{fig: contour of nCBF} is the slice of heatmap of $\hat{h}_w(x)$ with velocity $\dot{x}=0.2, \dot{y}=0.2$. }
\label{fig: 2D navigation}
\vspace{-20pt}
\end{figure*}

In this section, we use \gls{rl} to address the 2D navigation task introduced in Section~\ref{sec:2dNav}. Let $s_g = [x_g, y_g, 0, 0]$ be the goal state. The step reward is defined as $r_t = - 0.01 * \lVert s - s_g \rVert$, the terminal reward is $r_\mathrm{collision}=-5$ when the robot collides with the obstacles and $r_\mathrm{goal}=10$ when the robot reaches the goal area $X_g = \{s \mid |\lVert s - s_g \rVert < \epsilon \}$ where $\epsilon=0.1$ is the goal tolerance. We use Proximal Policy Optimization \cite{schulman2017proximal} to train the agent. To ensure safety during the training, we solve
\begin{equation}
    \begin{split}
        u_\mathrm{safe} = & \arg \min_{u \in \mathbb{U}_a} \| u - u_\mathrm{RL} \|^2 \\ 
        s.t. & \ L_f\hat{h}_w(x) + L_g\hat{h}_w(x)u + \alpha(\hat{h}_w(x)) \ge 0
    \end{split}
\end{equation} to project the action $u_\mathrm{RL}$ of the \gls{rl} policy to the safe action $u_\mathrm{safe}$ with the least modification.

We note that, theoretically, this controller guarantees safety with infinite control frequency. However, a continuous controller is not possible to implement on discrete control units. This limits the safety guarantees we may provide. How to address the gap between continuous controllers and their discrete implementations remains an open question. 

Figure \ref{fig: trajectories during training without nCBF} shows all trajectories performed during the training. We can see that several trajectories collide with the obstacles. Note, that the learned policy is not guaranteed to be safe. Figure \ref{fig: trajectories during training with nCBF} shows all the training trajectories with \gls{ncbf} as a safety filter and no trajectories are colliding with the obstacles. However, we observe that the average reward with \gls{ncbf} is larger than for nominal \gls{rl} without \gls{ncbf} in the very early stage but has a slower growth rate and converges to a lower reward level compared with nominal \gls{rl}, see Fig. \ref{fig: average reward during training}. The reason is that \gls{ncbf} provides prior knowledge about the environment and the agent could avoid exploring unsafe regions in the early stage and gain a higher reward than nominal \gls{rl}. However, the forward invariant set is still suboptimal as discussed in Section \ref{sec: Optimality}, which means only a suboptimal policy is learned and exploration is restricted. Nevertheless, we believe that provided safety guarantees are beneficial in safety-critical applications.
\section{Conclusion}
In this work, we presented a framework that simultaneously synthesizes and verifies continuous \acrfullpl{ncbf}. To this end, we leveraged bound propagation techniques and the \acrlong{bbs} to efficiently verify neural networks as \acrfullpl{cbf} in the continuous state space. 
In experiments, we showed that our framework efficiently synthesizes an \gls{ncbf} which renders a larger safe set than state-of-the-art methods without requiring an initial guess. 

Since the memory requirements and computation time of the \acrlong{bbv} increase exponentially with the system dimension, in future work, we may address the scalability of our framework.

\bibliography{references}
\bibliographystyle{ieeetr}

\end{document}